# Communications for Wearable Devices

Shivram Tabibu

Abstract. Wearable devices are transforming computing and the human-computer interaction and they are a primary means for motion recognition of reflexive systems. We review basic wearable deployments and their open wireless communications. An algorithm that uses accelerometer data to provide a control and communication signal is described. Challenges in the further deployment of wearable device in the field of body area network and biometric verification are discussed.

**Introduction**

Wearable wireless communication system is exciting new frontier techniques and it is adding another layer to human-machine interaction [1][2][3][4][5]. Wireless communication is necessary to make it possible for wearable devices to transmit data to proximate devices and this brings up many problems of transmission and software control 6][7][8]. There are many devices operating in the localized region or within human body contact, such as the smart phone watch, wearable computing devices, RFID and health care monitoring devices [9][10][11], and the RF band is shared with mobile/ cellphones, WLAN, Personal area networks, satellite communications and many other applications [12]-[21]. Data transmission must be secure since wearable devices transmit vital information and it is data that needs to be private and for this both access control and encryption must be employed [22]. Questions related to security over the network must also be addressed and future solutions that include quantum algorithms must be researched [23]-[30].

Several wireless communication systems deal with the specific issues of body centric communication using a device like a radio terminal placed on the skin surface of human body. The communication system aims at short-range communication like- personal area networks (PAN) and body area network (BAN), using human body as a transmission conduit.

The objective of this paper is to explore the usage of wearable devices for activity recognition by using wireless communication channel where the transmission of signal through body area network uses half duplex communication with the help of a wearable device wireless communication. The activity recognition shall be explained by using a single wearable device accelerometer sensor data to understand wireless communication mechanism of a single wearable device. The signal transmission network of a wearable device via human body or any reflexive system to deal with a propagating signal will also be examined.

**Background**

Prior research on body-centric communications explains the transmission process of a wearable device in terms of physical layer and interface of the electromagnetic (EM) wave with the human body. The electric-field distribution of a wearable device with human body as a transmission conduit was evaluated by the finite difference time domain (FDTD) method and confirmed via trial experiments [5].

The wireless communication system for a wearable device is illustrated by establishing an intra body network to transmit information to proximate devices. Figure 1 explains the hardware design of an Intra Body Communication System. The module comprises of a Serial interface unit (SIU), a serializer/deserializer and a modulator/ demodulator. In order to interface with embedded microcontrollers, the serial interface is used as an on-chip communication system.

The communication module is controlled by the integrated processor via serial interface. It transforms the data to serial form and applies FIFO to improve system efficiency. It is then modulated to turn into transmission signal for communication for Intra Body. Modulation along the signal transmitter is done by modulator where the serial data is modulated with a specified frequency in order to use a given system as a transmission line and there is a corresponding demodulator. Deserializer allows the transmission of data from demodulator to the main processor and the SIU. The system is implemented on FPGA with a size of wrist watch sized devices.

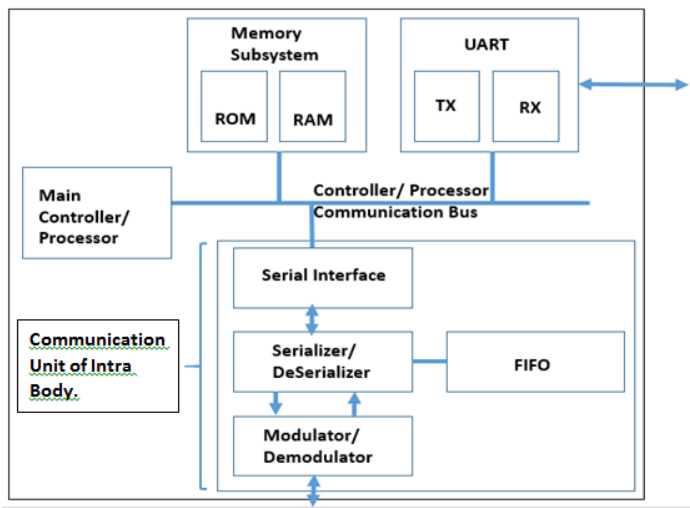

Figure 1. Block diagram- Hardware Architecture of a Communication System module of Intra Body [1].

Wireless communication may also be achieved via a wearable device (e.g. TI EZ 430 Chronos smart watch). The EZ 430-Chronos is a wireless wearable device with several activity recognition features. It comes along with an access point (i.e. the neighboring device the watch can communicate with) in the form of a dongle, communicating on one side to the watch through radio frequency wireless communication of 900 MHz and on the other side communicates with a PC by means of USB interface as shown in Figure 2.

The communication with the watch is to obtain graphical sensor data, perform sensor calibration and provide health data by connecting the USB interface with a main receiving/ processing station. The watch is basically a sensor node for wireless data reception. It is an integrated system consisting of a segmented LCD (96 segment display), a pressure sensor and three axis accelerometer sensor for detecting variations of different motion along with MSP 430 (a low-power microcontroller) development kit. The data acquired from 3-axis accelerometer sensor can be exploited to govern the electrical appliances in a home.

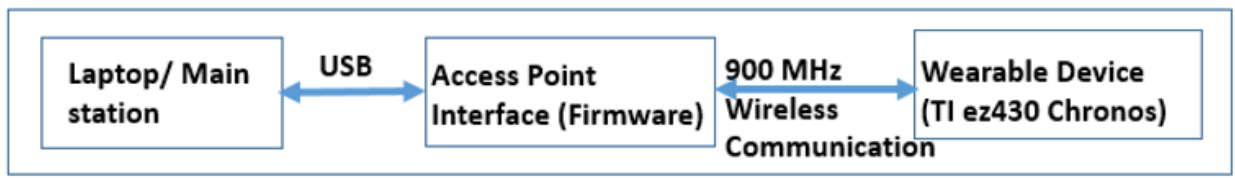

Figure 2. Block diagram of Hardware Architecture of an Intra Body Communication System/ module
(https://exosite.com/ti-chronos-ez430-watch-pc-gateway/)

The general RF communication is illustrated in figure 3. It is clear from the figure that an RF switch is integrated before the LNA module to increase the isolation between different communication bands. To increase the dual-band wireless receivers' sensitivity, LNA (Low Noise Amplifier) is used and it also prevents interference from undesirable signals.

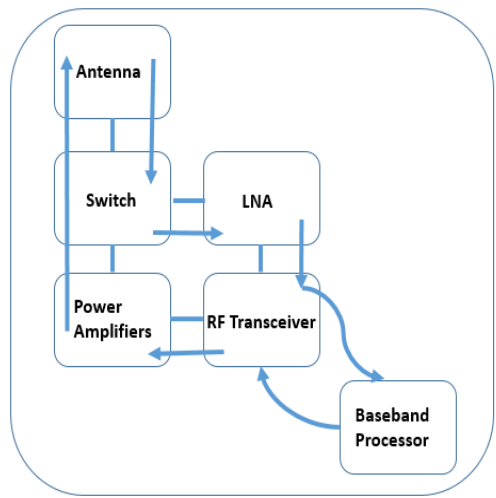

Figure 3. Block diagram of Hardware Architecture of an Intra Body Communication System/ module

## Wireless Communication for Home Automation using Wearable Device

The accelerometer integrated inside a wearable device is exploited in different ways to control various processes that may include appliances.

The primary idea is to transmit the attained human activity level by communicating through RF signals (Serial Communication) into a form of electrical energy which finally assists in controlling electrical appliances. Firstly, a person / human being is sited by the PIR sensor; then the watch checks the motion of the accelerometer and based on the data available a definite action is executed. The eZ430 is a CC430 microcontroller (Low Power RF Microcontroller) like the one used in sports for fitness tracking. One of its most unique features is that it can be used as a base platform for wearable systems or as wireless sensor node for data collection.

The software control center associated with eZ430- Chronos has numerous features validating the wireless communication competences of the watch on Windows and Linux operating system. Several RF protocol can be selected for use, depending on the purpose. The Chronos module communicates to PC through a RF USB debugging interface/ CC1111 RF USB access point. It transmits 3- axis acceleration values and after transmission is successful, the control center station reports as "Acquiring data from accelerometer sensor" and the values are displayed w.r.t each axis (X, Y, Z). It has these other features:

- 3 axis accelerometer sensor data graph
- PowerPoint/ remote control
- Synchronization of time, date and calibration data
- Heart beat rate
- Software update

The eZ430 consists of two important modes- ACC & PPT/ Synch mode. The watch has to be set to ACC mode in order to activate wireless communication of accelerometer data transmission. After the USB access point is started, the control center exhibits- "Access point started. Now start watch in ACC, PPT or Synch mode."

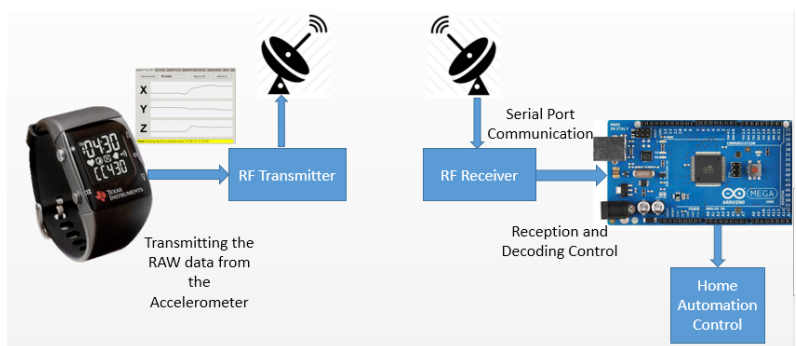

Figure 4. Block diagram of Hardware Architecture of an Intra Body Communication System/ module

The whole procedure involves the following steps:
- Transmission of the raw data from the TIEZ430 watch over the Radio Frequency
- Reception of the data and converting the data from the raw to required type
- Using this data for effective analysis and controlling the required electrical appliance

The communication starts with a PIR (Passive Infrared Sensor) sensor which is often used for motion detection in its vicinity. The specific motion of the watch is detected by the PIR sensor. When the particular PIR sensor is activated, it states that the human/ person has entered the room i.e., he/ she is in the locale of the sensor.

Once the particular motion of the watch is perceived by sensor, the light or the electrical appliance in the room gets powered ON. In addition to that, when the motion of the watch is activated again; the sensor identifies the signal and the light or the electrical appliance is turned OFF.

The wireless communication outline is illustrated in the following diagram:

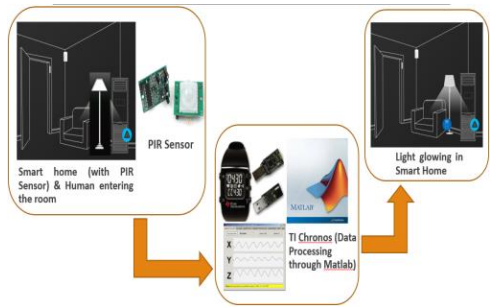

Figure 5. Wireless Communication outline structure using TI eZ430- Chronos for home automation

For example, if the motion of hand (with watch worn) is in vertical direction, the light gets turned ON. If the motion of the hand is horizontal, the light gets turned OFF. For other movements apartment from horizontal and vertical, there is no activity (DO NOTHING) in the room.

The algorithm to activate the watch is developed using Matlab and the completed data processing takes place within Matlab. The values obtained from the accelerometer sensor inside the watch are given below.

Table 1. Data obtained from the accelerometer sensor

| Hand-Wrist Movement | TI Chronos Accelerometer Values | Light/ Electrical Appliance Status |
|---|---|---|
| Motion Values (Z-Axis) Up-Down Movement of Hand/Wrist | 277 | ON |
| | 279 | ON |
| | 282 | ON |
| | 284 | ON |
| | 265 | ON |
| | 277 | ON |
| | 261 | ON |
| | 274 | ON |
| | 269 | ON |
| | 276 | ON |
| | 270 | ON |
| | 280 | ON |
| | 270 | ON |
| | 267 | ON |
| | 268 | ON |
| | 279 | ON |
| | 272 | ON |
| Motion Values (Y-Axis) Horizontal Movement of Hand/Wrist | 360 | OFF |
| | 363 | OFF |
| | 374 | OFF |
| | 379 | OFF |
| | 367 | OFF |
| | 326 | OFF |
| | 331 | OFF |
| | 356 | OFF |
| | 323 | OFF |
| | 381 | OFF |
| | 335 | OFF |
| | 359 | OFF |
| | 339 | OFF |
| | 368 | OFF |
| | 352 | OFF |
| | 378 | OFF |
| | 372 | OFF |
| | 335 | OFF |
| Other Movements | 230 | DO NOTHING |
| | 225 | DO NOTHING |
| | 228 | DO NOTHING |
| | 192 | DO NOTHING |
| | 219 | DO NOTHING |
| | 212 | DO NOTHING |
| | 217 | DO NOTHING |
| | 199 | DO NOTHING |
| | 208 | DO NOTHING |
| | 224 | DO NOTHING |
| | 211 | DO NOTHING |
| | 184 | DO NOTHING |
| | 182 | DO NOTHING |
| | 179 | DO NOTHING |
| | 184 | DO NOTHING |
| | 169 | DO NOTHING |
| | 201 | DO NOTHING |
| | 206 | DO NOTHING |
| | 215 | DO NOTHING |

The algorithm in the Matlab is written in such a way that first it calculates the mean average of all values for ON, OFF & DO NOTHING. Figure 6 presents the output of the RF USB access mode for the two case of "no signal" and of "trigger".

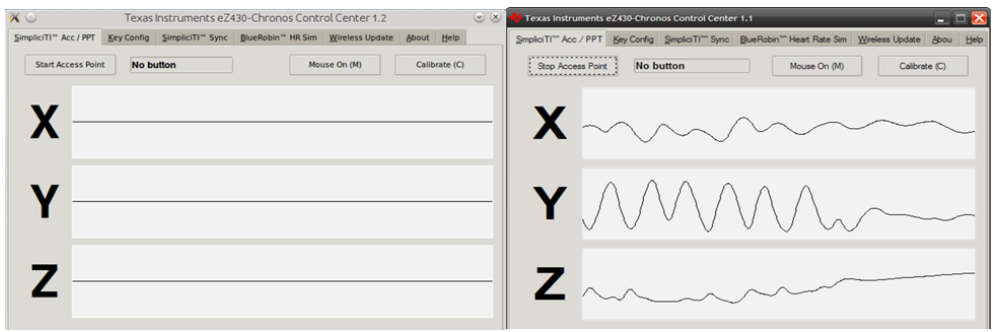

Figure 6. Off mode output (left); On mode output (right)

The flow diagram of Figure 7 describes the algorithm more clearly.

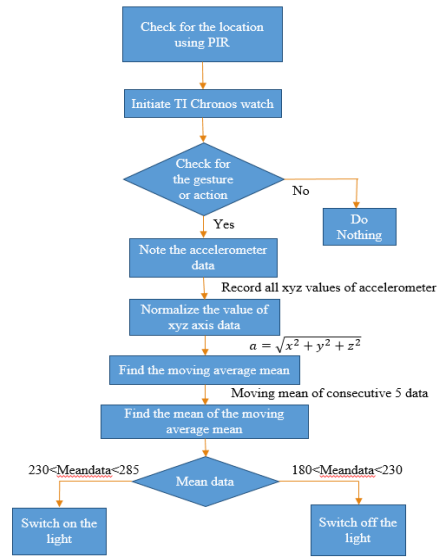

Figure 7. Working Flow diagram of the algorithm for processing TI eZ430- Chronos accelerometer data

## Working of the Algorithm

The computation is associated with PIR Sensor stimulation by sensing a human activity level in a particular room via TI Chronos smart watch. The following steps clarify the operation:

- Initially the watch is regulated through TI Chronos Control center software according to the human activity level because every human has different muscular reflexive action.

- Next, the TI watch is triggered to ACC mode before initiating data transmission.

- Subsequently the person wearing the wearable device enters a particular room where the PIR sensor is integrated. The corresponding PIR sensor gets triggered and it indicates the detection of person.

- After detection, the user triggers the watch in ACC mode and concurrently the Matlab algorithm is initiated to process the accelerometer data being continuously received from the watch.

- With respect to the motion, with in X/Y/Z axis, the appliance is triggered to ON state. In the experiment for turning the lights ON, we chose the hand motion across Z-axis resulting out the mean average value in the range of 240 to 286. For turning OFF the appliance (motion across Y-axis), the range of mean average value lies amid 323 to 384. For further movements apart from Y/Z axis, there is no intended action i.e., (DO NOTHING).

The following results show the output of the algorithm to ascertain that how a wireless communication via a small wearable device can be applied:

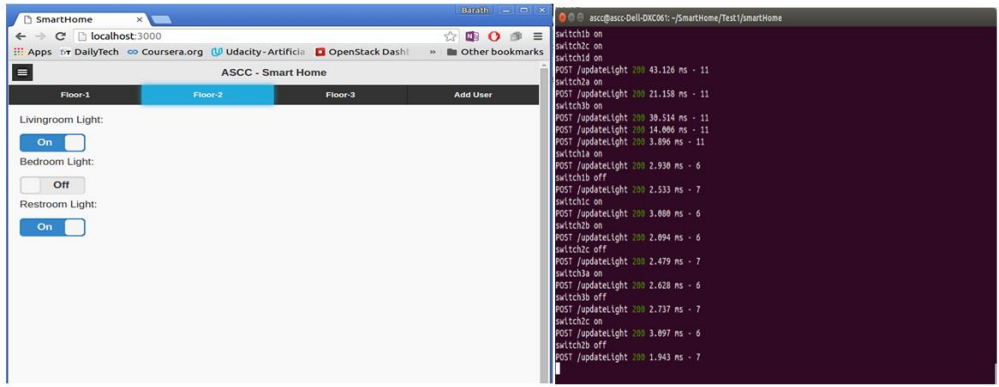

Figure 8. Output from the Home Automation Server for Wireless Communication via TI wireless device controlling home appliances

As it is clearly visible, that whenever the corresponding PIR sensor is activated using wireless device communication, the electrical appliance is either turned ON/ turned OFF via home automation wireless communication server.

## Applications

There ae several applications in several fields for wireless communication using wearable devices. Every medical, automotive, defense and manufacturing industry is trying to implement wearable devices for controlling the system, to optimize the power and make the process control flow easier. Some of the applications are as below: Home automation for motion controlled devices including home electrical systems/ appliances; Minimizing the work for elderly/ physically impaired people in functioning the home appliances; A methodology to control the excess consumption of electricity by switching the electrical appliance; Optimization of Power in process control industry; Automated missile technology; Wireless communication & tracking; Healthcare, monitoring & sensing; Smart cards, labels, devices.

Wearable devices can also be used in biometric applications that are completely dependent on the human body. Body sensor network/ body area network is a part of wireless communication for integrating several sensor nodes concentrated on a person's conduit. The prompt advancement of microprocessor technology in the field of wireless communication has advanced the technology of body area network in healthcare watching, blood/glucose level analysis, human sensing recognition, fitness monitoring etc. Since every wearable device usually contains personal information, hence data outflow from a wearable device is viewed as a big task, which may result in data loss. Thus, it is a primary task to ensure the security of data flow in wearable device.

This technology fulfils a current need in the field of information technology security. It uses human speech, eye-iris, finger-print or ECG/ EEG pattern for biometric verification. But it has several challenges that today's scientists are facing during experiments. Also the ECG/ EEG reports are not so reliable and EEG headset while the time of recording is sometimes difficult to wear for a longer duration. The iris verification is also challenging because of the limitations of wearable device build. Consequently, a novel methodology to biometric verification is needed in a wearable device.

## Conclusions

Current research is being carried out in this field in combining several wearable sensor devices with an accelerometer to recognize daily activities and transmit the information to the receiving station. The governing expert recommends the small spectrum of RF band for technical study and further research. The exploration on hardware resources and life of battery required for wearable device to establish a long term wireless communication channel has not achieved a breakthrough yet. In addition to it, the key aim for the communication for wearable devices includes the following:

- Minimize the number of iterations essential to trigger the electrical device
- Judgement/ decision making in a less amount of time
- Reduce the number of times the accelerometer is reset

With the transformation of wireless communication in every field/ industry, it is expected that wearable devices will be available to the general public. It is not unreasonable to assume that several medical examinations, industrial power optimization and process flow control requiring a person to trigger the corresponding activity shall eventually be performed by wearable devices.

Wearable devices shall become the mainstream for advancement of mobile smart devices which in the long run shall change completely the modern way of life. Using a wearable device sensor, several other daily activities could be recognized by combining pressure sensors & accelerometers and extracting data via wireless communication for further analysis.

Wearable device technology combined with big data analysis provides a pathway to another revolution. Even in the big data milieu via wireless communication, the wearable market shall have a greater impact & development creating huge capital along with making people's life more opportune.